\documentclass[prb,twocolumn,aps,showpacs]{revtex4}
\usepackage{graphicx,color}
\usepackage{latexsym,amssymb}

\def\be{\begin{equation}}
\def\ee{\end{equation}}

\def\bc{\begin{center}}
\def\ec{\end{center}}

\begin{document}

\title{Disorder-driven splitting of the conductance peak at the Dirac point in
graphene}

\author{L. Schweitzer$^{1}$ and P. Marko\v{s}$^2$}
\affiliation{%
$^1$Physikalisch-Technische Bundesanstalt (PTB), Bundesallee 100, 38116 Braunschweig, Germany\\
$^2$Department of Physics, FEI, STU, 812\,99 Bratislava, Slovakia}

\begin{abstract}
The electronic properties of a bricklayer model, which shares the same topology
as the hexagonal lattice of graphene, are investigated numerically. We study the
influence of random magnetic-field disorder in addition to a strong
perpendicular magnetic field. We found a disorder-driven splitting of the
longitudinal conductance peak within the narrow lowest Landau band near the
Dirac point. The energy splitting follows a relation which is proportional to
the square root of the magnetic field and linear in the disorder strength. We
calculate the scale invariant peaks of the two-terminal conductance and obtain
the critical exponents as well as the multifractal properties of the chiral
and quantum Hall states. We found approximate values $\nu\approx 2.5$ for the
quantum Hall states, but $\nu=0.33\pm 0.1$ for the divergence of the correlation length
of the chiral state at $E=0$ in the presence of a strong magnetic
field. Within the central $n=0$ Landau band, the multifractal properties of
both the chiral and the split quantum Hall states are the same, showing a
parabolic $f[\alpha(s)]$ distribution with $\alpha(0)=2.27\pm 0.02$. In the
absence of the constant magnetic field, the chiral critical state is determined
by $\alpha(0)=2.14\pm 0.02$. 

\end{abstract}


\maketitle

\section{Introduction}
The nature of the current carrying states near the charge-neutral Dirac point
in graphene is of exceptional interest and substantial importance for the
understanding of the electrical transport properties in strong magnetic fields.
The experimental observation of quantum Hall 
plateaus\cite{Nea05,ZTSK05,Zea06,Nea07,GZKPGM07,JZSK07} 
with $\sigma_{xy}=(2N+1)\,2e^2/h, N=0,1,\ldots$
strikingly emphasizes the significance of disorder, which influences the single 
sheet of carbon atoms forming the  hexagonal lattice of graphene. Due to the 
two valleys appearing in the band structure, each Landau band contributes
two times $2e^2/h$ to the Hall conductivity $\sigma_{xy}$, whereas the factor 
of 2 accounts for the spin degeneracy. 
  
In the absence of a magnetic field, simple on-site (diagonal) disorder gives
rise to strong Anderson localization, which causes the electrical current to
vanish at zero temperatures.\cite{AE06,Alt06} This is, however, in
conflict with experimental observations\cite{Nea05,ZTSK05,Mea06,Zea06,JZSK07}
demonstrating that either a different type of disorder is present in real
samples or that electron-electron interaction renders the one-particle picture
obsolete.  
A disorder type being able to account for a finite conductance, is the 
ripple-disorder\cite{Mea06,Mea07} which is believed\cite{MG06,GMV08} to create
similar effects as those originating from fluctuations of the hopping terms 
due to elastic strains of the intrinsic curvatures of the graphene sheet.
It is also well known that the inter-valley scattering depends crucially on the
type of disorder.\cite{OGM06,NM07,OGM07,OGM07a,Khv08}

Recently, a splitting of the conductivity maximum within the central Landau
band at the Dirac point has been observed in high mobility graphene samples
for very strong magnetic-flux densities $B > 20$\,T.\cite{Zea06,JZSK07,Aea07,GZKPGM07} 
The measured energy splitting $\Delta E \propto\sqrt{B}$ 
has been suggested\cite{JZSK07} to be due to a lifting of the
sub-lattice symmetry caused by electron-electron interaction. 
Also, the effect of counter-propagating chiral edge states\cite{Aea07}, 
electron-lattice effects,\cite{FL07} as well as valley
ferromagnetism\cite{NM06,ALL06,FB06,GMD06,AF06} have been put forward to 
account for the observed splitting near the Dirac point.  
Although the proposed approaches contain many interesting physics
based on interaction effects, we would like to retain the non-interacting
particle picture in the present work and investigate the influence of a random
magnetic field (RMF) which causes similar effects as 
ripple-disorder.\cite{MG06,GMV08}
Recently, the influence of real random hopping terms was studied 
and a splitting of the extended state in the $n=0$ Landau band has been
reported.\cite{KA07} However, the experimentally observed extremely narrow
Landau band\cite{Zea06,JZSK07,Aea07,GZKPGM07} at the Dirac point suggests the
origin of the Landau-level broadening to be ripple-disorder or an equivalent
disorder that also preserves the chiral symmetry.\cite{OGM08}     

By using of a microscopic bricklayer model, which is topologically
equivalent\cite{HHKM90,WFAS99} to a hexagonal lattice, and assuming a random
magnetic-flux disorder by introducing complex phases into the hopping terms of
a tight-binding Hamiltonian, we found a narrow density-of-states peak at the
Dirac point and a splitting of the central conductance peak 
similar to what has been observed in experiments.  We calculate the density of
states, the two-terminal conductance, and the critical eigenstates, from which
the respective energy and magnetic-field dependence of the energy splitting,
the scaling of the conductance, and the multifractal properties of the
critical eigenstates are obtained. The splitting $\Delta E$ increases linearly
with the strength of the random flux amplitude and shows a $\sqrt{B}$
dependence as observed in experiments. Besides the split quantum Hall
conductance peak, we found a central chiral state at $E=0$.  The latter
exhibits a critical exponent $\nu=0.33 \pm 0.1$, which determines the
divergence of the localization length $\xi/\xi_0=|E|^{-\nu}$, whereas the
remaining two split bands of the central Landau level belong to the ordinary
quantum Hall symmetry class with a critical exponent $\nu\approx 2.5$. 
The calculated multifractal properties turn out to depend on the static
magnetic field. For $B=0$, we found $\alpha(0)=2.14$ for the chiral state at 
the Dirac point. This value changes to the usual quantum Hall result
$\alpha(0)=2.27$ in the presence of a spatially constant magnetic field.

\section{Bricklayer model and transfer matrix method}

\begin{figure}
\includegraphics[width=8.0cm,clip]{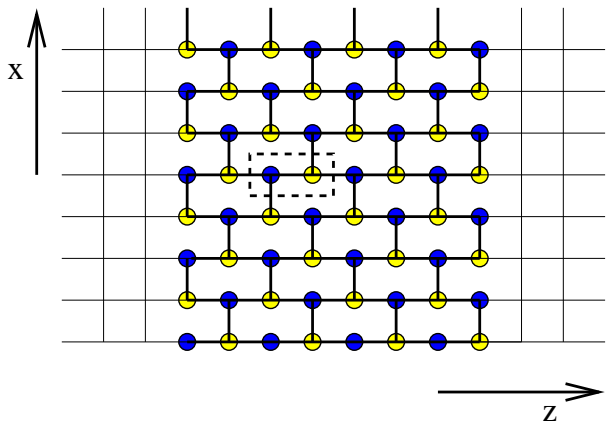}
\caption{(Color online)  
The two-dimensional bricklayer lattice which shares the same topology as the
hexagonal lattice of graphene. The bi-atomic unit cell is indicated by the
dashed rectangle. For the calculation of the conductance, the sample is
connected to two semi-infinite leads (thin lines) with square lattice
topology, and periodic boundary conditions are applied in the vertical direction 
(top vertical lines are connected to sites in the lowest horizontal line). 
}
\label{bricklayer}
\end{figure}

Graphene can be represented by a tight-binding Hamiltonian defined on a 
two-dimensional honeycomb lattice. In our numerical calculations, the
honeycomb lattice is transformed into a bricklayer lattice as shown in
Fig.~\ref{bricklayer}. Each site is connected by three bonds with its nearest
neighbors and has the same topology as the hexagonal
lattice.\cite{HHKM90,WFAS99} For investigations of the spectral properties,
the differences in bond length and bond angle do not matter. In 
other cases the length scale has to be properly adjusted, e.g., for the
plaquette size $2a^2$ (bricklayer) and $(3\sqrt{3}/2)a^2$ (hexagonal
lattice), where $a$ is the respective nearest-neighbor distance.

A magnetic field perpendicular to the $xz$ plane creates a magnetic flux 
through each individual plaquette,
\be\label{flux}
\Phi_{x,z} = \frac{p}{q} h/e + \phi_{x,z}
\ee
where $p$ and $q$ are integers which are mutual prime, and $\phi_{x,z}$ is the
random flux component. The latter is assumed to be due to ripples and to the
buckling of the non-planar carbon monolayer which causes the magnetic flux to
fluctuate from plaquette to plaquette. Since the plaquette size on the bricklayer 
system is twice that of the square lattice, the magnetic-flux density is
$B=p/q\times h/(e2a^2)$. We choose $\phi_{x,z}$ to be uniformly distributed 
according to $-f/2 \le \phi_{x,z}/(h/e) \le f/2$ so that the mean value $\langle
\phi_{xz}\rangle$ is zero and the variance is $f^2/12$. 
The parameter $f$ measures the strength of the disorder with a maximal value
$f=1$. 

The corresponding tight-binding Hamiltonian, with spinless fermionic particle
creation ($c^{\dagger}$) and annihilation ($c$) operators, on the bricklayer is
\begin{eqnarray}
{\cal H}  &=& 
V\sum_{x,z}{}' e^{i\theta_{x,z}}c_{x,z}^\dag c_{x+a,z} +
e^{-i\theta_{x-a,z}}c_{x,z}^\dag c_{x-a,z} 
\nonumber \\
& &  
+ V\sum_{x,z} c_{x,z}^\dag c_{x,z+a}+  c_{x,z}^\dag c_{x,z-a}.
\label{ham}
\end{eqnarray}
The complex phase factors along the vertical bonds of a given plaquette are
determined by the respective flux, $\theta_{x,z+2a}-\theta_{x,z}=2\pi\Phi_{x,z}$. 
We fix the length scale by $a=1$ and set $V=1$, which defines the energy scale. 
Please note that the first sum $\sum'$ in Eq.~(\ref{ham}) contains only the
non-zero vertical hopping terms as shown in Fig.~\ref{bricklayer}. 
An important property of the random flux model is that its disorder does not
destroy the chiral symmetry of the system at the band center 
$E=0$.\cite{MW96,Fur99,MBF99,BMSA98} Thus, the 
Hamiltonian (\ref{ham}) enables us to study both the chiral and the quantum
Hall critical regimes within the same model.

For the calculation of the conductance, the samples are attached to two
semi-infinite leads, which are constructed by a two-dimensional square 
lattice. No magnetic field is considered within the two leads. Periodic
boundary conditions are assumed in the vertical ($x$) direction in order to
eliminate surface effects and edge states in the quantum Hall regime. This
requires the number of sites in the vertical direction to be even, and a
multiple of $2q$. 

The two-terminal conductance is calculated by the well-known transfer-matrix
method\cite{PMR92} 
\be
g=\textrm{Tr}~t^\dag t=\sum_i^{N_{\rm ch}}\frac{1}{\cosh^2(\epsilon_i/2)},
\ee
where $N_{\rm ch}$ is the number of open channels, $t$ is the
transmission matrix, and the $\epsilon_i$ parametrize its eigenvalues. 
Statistical ensembles of typically $N_{\rm stat}=10^4$ samples were collected 
and the mean value of the conductance was calculated for each set of parameters
($E,L,f, p/q$). The transformation of the honeycomb lattice into the bricklayer
changes the length scale along the propagation direction, which is due to the
rectangular shape of the unit cell. For computational efficiency, we still
consider $L\times L$ bricklayer lattices. We verified that for different
shapes the conductance scales like $g_c\sim L_x/L_z$ at all critical points. 
Hence the conductance results we show are always by a factor of two
larger. Therefore, it is easy to correct for our special aspect ratio when
comparing with conductance results obtained by other researchers.  

\begin{figure}
\includegraphics[width=8.0cm]{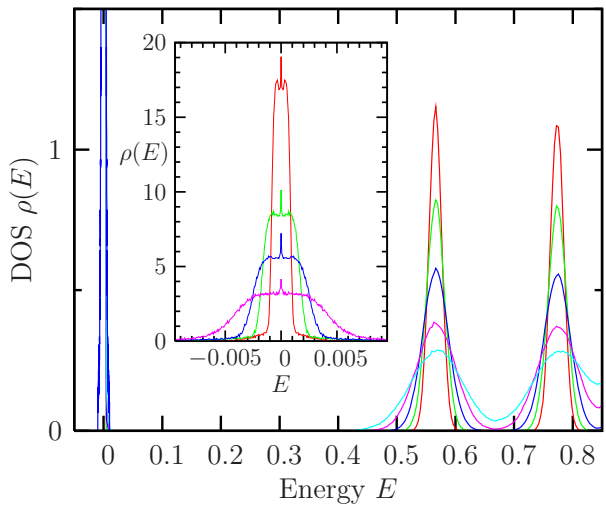}
\caption[]{(Color online) The density of states for $L\times L$ bricklayer 
  samples with size $L=128$, averaged over 500 disorder realizations, showing
  three Landau bands closest to the Dirac point for energy $E\gtrsim 0$. The
  random flux disorder strengths are $f=0.001, 0.002, 0.005, 0.007,
  \textnormal{ and } 0.015$, and $p/q=1/32$. The inset shows the
  broadening of the narrow central Landau band together with the chiral state
  at $E=0$  for $f=0.001, 0.002, 0.003, \textnormal{ and } 0.005$.}
\label{figdos}
\end{figure}

\section{Results and Discussion}
\subsection{Density of states}
The density of states $\rho(E)$ was calculated by counting the eigenvalues
obtained from diagonalization of $L^2=128\times 128$ samples with periodic
boundary conditions in both directions, averaged over 500 disorder
realizations. Without disorder, $\rho(E)$ was checked to be identical with the
result of a true honeycomb lattice.  Figure~\ref{figdos} shows 
$\rho(E)$ for $p/q=1/32$ flux quantum and several disorder strengths
$f=0.001, 0.002, 0.003, \textnormal{and\ } 0.005$. Due to the symmetry around
$E=0$, the density of states is shown only for energies $E\gtrsim 0$. The Landau
spectrum and the broadening of the bands are visible. However, compared with
the $n=1$ and $n=2$ Landau bands, the central $n=0$ one remains very narrow
and is hardly discernible. Therefore, in the inset of Fig.~\ref{figdos}, 
the broadening of the
narrow central band is shown in more detail where also the chiral peak at
$E=0$ is distinguishable in the middle of the broadened Landau band. A real
splitting of the $n=0$ Landau band becomes apparent only for very small
disorder strength $f$. The total width at half height of the central band
(neglecting the narrow chiral peak) is proportional to $\sqrt{p/q} f$. 
It should be noted that the degeneracy of the $n=0$ Landau level 
found in the Dirac model, even in the presence of a random magnetic
field,\cite{AC79} is already lifted in our clean lattice model due to Harper's 
broadening.\cite{Har55} In the case $B=0$, this lattice effect also causes
deviations from the linear energy dispersion away from the Dirac point.   

\begin{figure}
\includegraphics[width=8.5cm,clip]{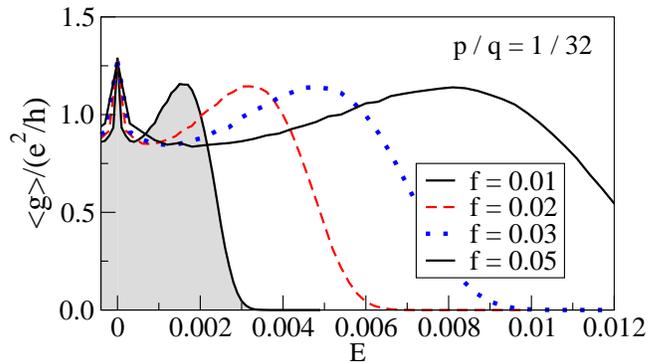}
\caption[]{(Color online) The energy dependence of the disorder averaged
  two-terminal conductance $\langle g(E)\rangle$ for $E\ge 0$, $p/q=1/32$,
  and various random flux disorder strengths $f$. The size of the square
  samples is $L=128$. Besides the conductance peak of the chiral state, which
  remains at $E=0$, the shift of the split quantum Hall state with disorder is
  depicted.}  
\label{fig_g_E}
\end{figure}

Since the number of eigenvalues is the same for each principle band shown in
Fig.~\ref{figdos}, 
the two split $n=0$ sub-bands must contain less eigenvalues than the higher
Landau bands due to the additional chiral states close to the Dirac point
$E=0$.  

\subsection{Conductance peak splitting}
\begin{figure}
\includegraphics[width=8.0cm]{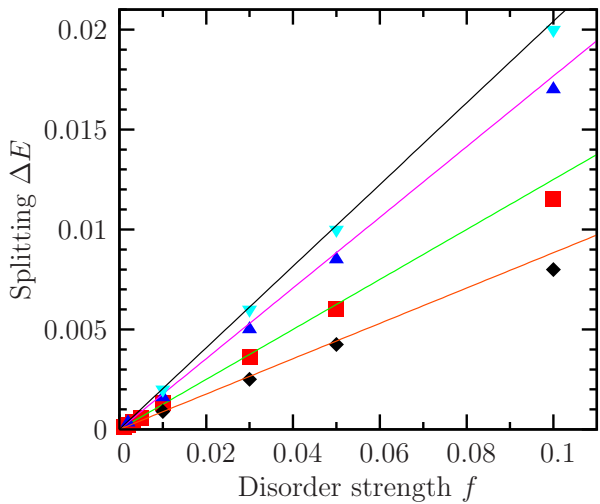}
\caption[]{(Color online) The splitting of the conducting states from the
  $n=0$ Landau band obtained from the two-terminal conductance of $L=256$
  square samples versus random magnetic field disorder strengths $f$. The
  magnetic fields are $p/q=1/24$ ($\blacktriangledown$), $p/q=1/32$
  ($\blacktriangle$), $p/q=1/64$ ($\blacksquare$), and $p/q=1/128$
  ($\blacklozenge$), respectively. The straight lines follow $\Delta
  E=\sqrt{p/q}\,f$.}    
\label{fig_splitt}
\end{figure}

The energy dependence of the disorder averaged two-terminal conductance is
shown in Fig.~\ref{fig_g_E} for different random magnetic-field disorder
strengths $f$. Again, only the positive energies are shown, because the
conductance is an even function of the energy $E$. With increasing $f$, the
conductance peaks corresponding to the $n=0$ Landau level, with peak value of
about $1.12\,e^2/h$, move away from $E=0$, where the conductance peak 
of the scale independent chiral state remains fixed at $g(E=0)=1.27\,e^2/h$. 
The disorder induced splitting $\Delta E$ of the $n=0$ conductance peak is
plotted in  Fig.~\ref{fig_splitt} versus $f$ for various magnetic fields with
$p/q=1/24, 1/32, 1/64, \textnormal{ and }1/128$. For not too large $f$, the
data points follow the straight solid lines which are given by the relation 
\begin{equation}
\Delta E=\sqrt{\frac{p}{q}} f.
\end{equation}
Therefore, the conductance peak splitting shows a $\sqrt{B}$ behavior as
observed in experiments and increases linearly with the strength $f$ of the
random magnetic-flux disorder. Neither a splitting nor any shift of the
conductance peak due to the disorder $f$ investigated could be
observed for the $n=1$ Landau band and $L\le 256$.    

\begin{figure}
\includegraphics[width=8.0cm]{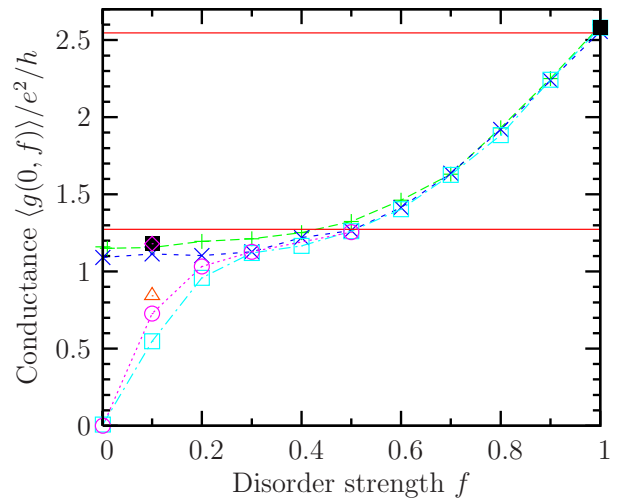}
\caption[]{(Color online) The averaged two-terminal conductance $\langle
  g(E=0,f)\rangle$ for several magnetic fields (given by the ratio $p/q$)
  versus random magnetic-field disorder strengths $f$. $p/q=1/64$: $L=256$ (+),
  $L=512$ ($\Diamond$), $L=896$ ($\blacksquare$). $p/q=1/128$: $L=256$
  ($\times$). $B=0$: $L=256$ ($\Box$), $L=512$ ({\Large $\circ$}), and $L=896$
  ($\triangle$). The horizontal lines mark the conductance values
  $4/\pi\,(e^2/h)$ and  $8/\pi\,(e^2/h)$, respectively.}    
\label{fig_g_0_h}
\end{figure}

As known from experiment,\cite{Tea07} the minimum conductivity at the Dirac
point depends verifiably on disorder. Therefore, in 
Fig.~\ref{fig_g_0_h} we show the disorder dependence of the critical  
chiral conductance $\langle g(E=0,f)\rangle$ for different magnetic fields. 
The two values $4/\pi\,(e^2/h)$ and $8/\pi\,(e^2/h)$ are indicated by solid
horizontal lines. 
The conductance turns out to be scale invariant for sufficiently strong random
flux disorder.  In the limit of $f\to 1.0$, $\langle
g(E=0)\rangle$ converges to a common value that is a little larger than
$8/\pi\,(e^2/h)$, which is independent of magnetic field. 
In contrast, a magnetic field dependence is seen for small $f$. In the range
$0.001 < f < 0.4$, the conductance seems to approach a value somewhat smaller
than $4/\pi\,(e^2/h)$ for all magnetic fields $p/q<1/32$ studied. 
For small disorder, e.g., $f=0.1$ and $B=0$, the conductance scales with the
sample size. A convergence to a value $\approx 4/\pi\,(e^2/h)$ as in the
finite $B$ case is compatible with our data.

In a finite clean system, $\langle g(E=0,f=0)\rangle$ is zero for $B=0$
but increases with the system size for finite $B$. The former behavior is due
to the vanishing density of states at $E=0$ in the absence of a magnetic
field. In the thermodynamic limit, the result for the conductance seems to
depend on the order of limits $L\to\infty$ and $f\to 0$. A special case
appears in the clean limit if $L_z$ is a multiple of three. $\langle
g(E=0,f=0)\rangle$ turns out to be $2e^2/h$ due to the four eigenstates
appearing at $E=0$ for $k_z = \pi/3$, where $k_z=N \pi/L_z$ and
$N=1,2,\ldots$\ . This agrees well with the experimentally observed minimal
conductivity $\sigma_{\rm min}=4e^2/h$ if the electron spin is taken
into account. 
   
\subsection{Critical quantum Hall regime}
The energetic position of the split conductance peak within the $n=0$
Landau band as shown in Fig.~\ref{fig_g_E} was estimated from the energy 
and size dependences of the conductance peak. For all values of disorder
strength $f$, the conductance $g(E)$ exhibits a maximum around the critical
energy $E_c(f)$. The maximal value of the conductance is $g_c\approx
1.12\,e^2/h$.  If we take into account the factor of two due to the special
aspect ratio, this value is in agreement with the critical conductance  
$g_c=0.6\,e^2/h$ found on the square lattice with spatial correlated diagonal
disorder.\cite{SM05} Likewise, it compares well with the results 
$g_c=2/\pi\,e^2/h$ obtained within a self-consistent Born approximation
by Shon and Ando\cite{SA98} for a graphene sheet in the presence of short-
or long-range scattering potentials. 
Finally, our result for the split conductance peak agrees also with the analytical
calculations of Ostrovsky and co-workers.\cite{OGM06,OGM07,OGM07a}

\begin{figure}
\includegraphics[width=7.5cm,clip]{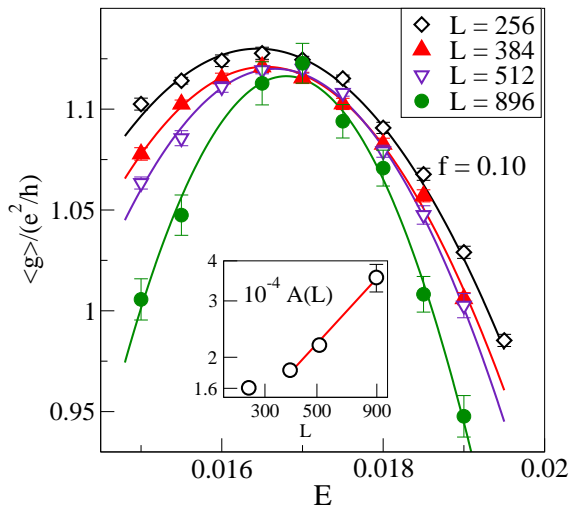}
\caption[]{(Color online) The energy dependence of the averaged conductance
  $\langle g\rangle$ for $p/q=1/32$ and $f=0.10$ calculated for the bricklayer
  model of size  $L\times L$. Solid lines are quadratic fits, given by
  Eq.~(\ref{quad}). A scaling analysis provides us with a
  critical exponent $\nu\approx 2.5$ if only the data for $L\ge 384$ were used.
  The inset shows a double-logarithmic plot of the size dependence of $A(L)$
  defined in Eq.~(\ref{quad}). The solid line is a power-law fit of
  Eq.~(\ref{al}) with $\nu\approx 2.5$.
}
\label{brick_32_f010}
\end{figure}

A more detailed analysis of the size and energy dependences of the conductance 
shows that $g(E)$ exhibits a pronounced maximum around the disorder dependent
critical energy $E_c(f)$. This renders the estimation of the corresponding
critical exponent possible.  
The peak $g(E,f)$ becomes narrower when the size $L$ of the system increases.
Following the scaling theory of localization, the conductance is assumed to be
a function of only one parameter in the vicinity of the critical point, 
$\langle g\rangle = F[L/\xi(E)]$, with a diverging correlation length
$\xi(E)\propto |E-E_c|^{-\nu}$.  Then, we can extract the critical value of
the conductance and the critical exponent assuming that
\be\label{quad}
\langle g\rangle = g_c - A(L)\left[E-E_c\right]^2.
\ee
and
\be\label{al}
A(L)\propto L^{2/\nu}.
\ee
Figure~\ref{brick_32_f010} shows the size and energy dependence of the
conductance peak for disorder $f=0.1$. We obtained a critical exponent
$\nu\approx 2.5$ when only the data for $L\ge 384$ were used in the scaling
analysis.  
We observed a similar behavior for smaller disorder strength $f=0.01$.
In view of the uncertainties, we consider this to be consistent with the
well-known results $\nu= 2.3$ for ordinary quantum Hall systems supporting the
view of the universality of the continuous quantum Hall phase transition. 

\begin{figure}
\includegraphics[clip,width=7.5cm]{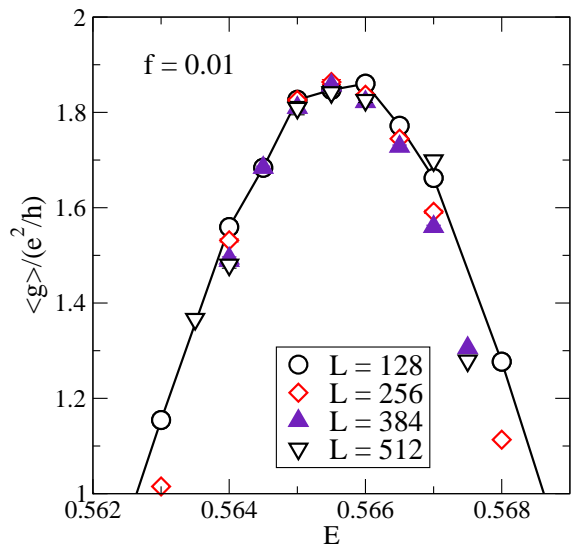}
\caption{(Color online) The disorder averaged two-terminal conductance versus
  energy in the $n=1$ Landau band for various system sizes $L=128$, 256, 384,
  and 512. The magnetic field is $p/q=1/32$ and the disorder strength $f=0.01$. 
}
\label{g_E_L_2nd}
\end{figure}

The two-terminal conductance for the $n=1$ Landau band is shown in
Fig.~\ref{g_E_L_2nd} versus energy. The peak value is about $1.87\,e^2/h$.
The maximal available size of the system ($L=512$) is still insufficient for a  
more accurate estimation of the critical parameters.
Therefore, we can neither give any limiting value nor address the Landau-level
dependence of the critical conductance. 
The very slow $L$ dependence of the conductance in the critical region,
indicating a larger localization length than for the $n=0$ case, makes the 
scaling analysis and estimation of critical exponent impossible.

\subsection{Critical chiral state}
\begin{figure}
\includegraphics[clip,width=7.5cm]{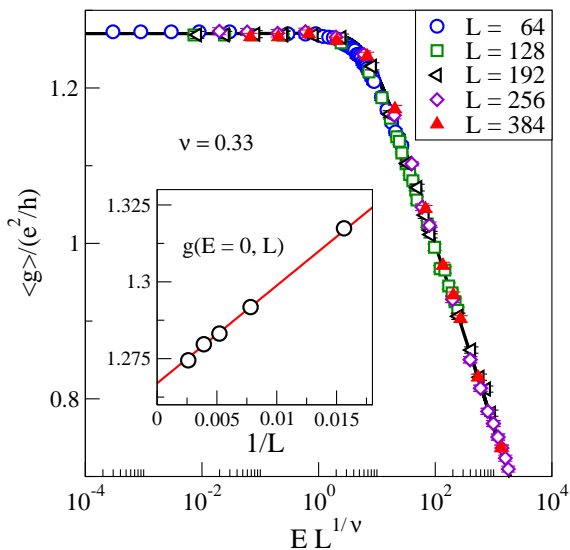}
\caption{(Color online) Scaling of the averaged chiral conductance in the
  vicinity of the critical Dirac point $E=0$. The magnetic field is
  $p/q=1/32$ and the disorder strength $f=0.01$. The data show scaling 
  with a critical exponent $\nu=0.33\pm 0.1$. 
  The solid line is the fitted function (\ref{corrfunct}).
  The inset shows the finite-size correction $g(E=0,L)/(e^2/h) = 1.267 + 3.178/L$, 
  and var~$g\approx 0.23\,(e^2/h)^2$. 
}
\label{chiral}
\end{figure}

From the energy and size dependences of the mean conductance in the vicinity of
the chiral critical point we extract the critical conductance,
$g_c=\lim_{L\to\infty} \langle g(E=0,L)\rangle$.    
Our data for the mean conductance $\langle g\rangle$ at the band center
$E=0$ exhibit a convergence to the size independent limit $g_c$,
\be\label{fisicorr}
\langle g(E=0, L)\rangle = g_c + b/L.
\ee
As shown in the inset of Fig.~\ref{chiral}, $g_c$ $\simeq 1.267\,e^2/h$. This
perfectly agrees with the theoretically predicted value $4/\pi\,e^2/h =
1.273\,e^2/h$ for the spin-degenerate situation.\cite{OGM06,OGM07,OGM07a} 
We mention again that in our case the factor of two originates not from the
spin but from the special aspect ratio used.

\begin{figure}
\includegraphics[clip,width=7.25cm]{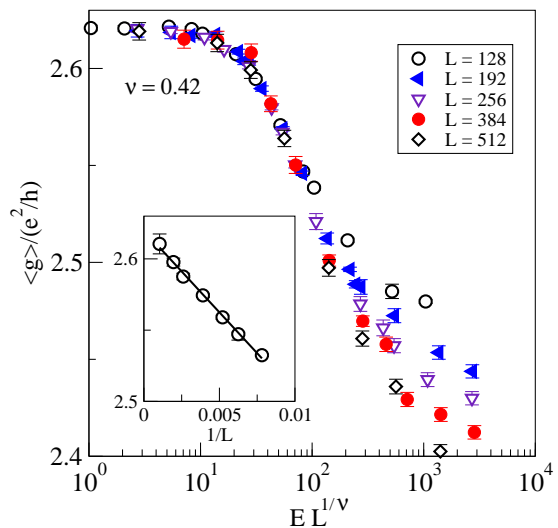}
\caption{(Color online) Attempt to scale the averaged conductance data close
  to the chiral 
  critical point for a  bricklayer system with zero homogeneous magnetic
  field and randomness $f=1.0$. Bricklayer square samples of size $L=128$,
  192, 256, 384, and 512 were considered. We obtain a critical exponent
  $\nu=0.42$. The inset shows the size dependence of $g(E=0)$ for $128\le L\le
  960$. The solid line is $g(L)/(e^2/h) = 2.62 - 11.48/L$, and var~$g\approx
  0.21\,(e^2/h)^2$.   
}
\label{chiral-B0}
\end{figure}

To estimate the conductance fluctuations, we calculated the probability
distribution $p(g)$ which proved to be Gaussian, confirming the
presence of mesoscopic fluctuations also at the chiral critical point. 
The variance var~$g = \langle g^2\rangle - \langle g\rangle^2$  varies 
slightly as a function of $B$ and $f$. For instance, we found var~$g\approx
0.23\,(e^2/h)^2$  for the data shown in Fig.~\ref{chiral} but
$0.21\,(e^2/h)^2$ for the $B=0$ case (Fig.~\ref{chiral-B0}).  

To analyze the critical behavior of the conductance in the vicinity of the
band center, we calculated $\langle g\rangle$ for square samples of size
$L=64$, 128, 192, 256, and 384, within the 
narrow energy interval $10^{-9}\le E\le 10^{-4}$. Our data confirm
that, after finite-size correction (\ref{fisicorr}), the mean conductance 
is a function of one parameter only, $\langle g(E,L)\rangle = G(EL^{1/\nu})$.
Following scaling theory, $\nu$ is the critical exponent that  governs the
divergence of the correlation length, $\xi(E)\propto |E|^{-\nu}$.
A more detailed analysis shows that 
\be\label{corrfunct}
G(x)= g_c -c_0\ln(1+c_1x+c_2 x^2),
\ee
with $c_0=0.05$, $c_1=-0.02944$, and $c_2=0.02134$. Scaling analysis provides us
with the critical exponent $\nu=0.33\pm 0.1$. Although the best scaling is
obtained for $\nu=0.33$ (as shown in Fig.~\ref{chiral}), a reasonable scaling
is also possible for values of the critical exponent in the range $0.25 < \nu
< 0.45$. Again, only numerical data for much larger systems would reduce this
inaccuracy.  

We have carefully checked that our estimate of $\nu$ is insensitive
to the range of the energy interval selected. Also, we tried to 
compensate for the observed divergence of the critical density of states near
$E=0$, which, however, changed our estimates only marginally. 
We mention that our result does not satisfy the Harris criterion,\cite{Har74}
which states that $d\nu-2\ge 0$. There are similar results also for other
models\cite{MH98,Cer00,ERS01,Cer01,MS07,SM08} with chiral critical exponents
$\nu<1$.   

In the absence of a constant magnetic field, we observed that the value of the
conductance at $E=0$ depends considerably on the size of the system,
$g(L)=2.62 - 11.48/L$. Even worse, the scaling behavior is fulfilled only in
a very narrow interval of conductance values. From Fig.~\ref{chiral-B0} we
see that the width of the scaling interval is comparable with the uncertainty
in the conductance due to the finite-size effects. It is therefore no surprise
that the estimation of the critical exponent is much more difficult.
Taking into account these restrictions, we conclude that the estimated value
of the critical exponent $\nu\approx 0.42$ is still in reasonable agreement
with the previously obtained values $\nu=0.35$ and $\nu=0.42$ for a normal
square lattice.\cite{MS07,SM08}  

\subsection{Multifractal eigenstates} 
The critical eigenfunctions $\psi_E(r)$ of bricklayer samples with sizes
up to $L^2=384\times 384$ were obtained numerically using a Lanczos
algorithm. For eigenvalues close to $E=0$, the chiral eigenstates show the
expected sub-lattice polarization, since for a clean system the wave functions
are non-zero only on one of the sub-lattices.\cite{KA07,PS08} 
This sublattice polarization is still observable in the presence of both the
fluctuating random flux and the constant perpendicular magnetic field. 
We also notice some weak quasi-one-dimensional structures making the appearance of
$|\psi_{E=0}(r)|^2$ look slightly anisotropic. 

\begin{figure}
\includegraphics[clip,width=8.7cm]{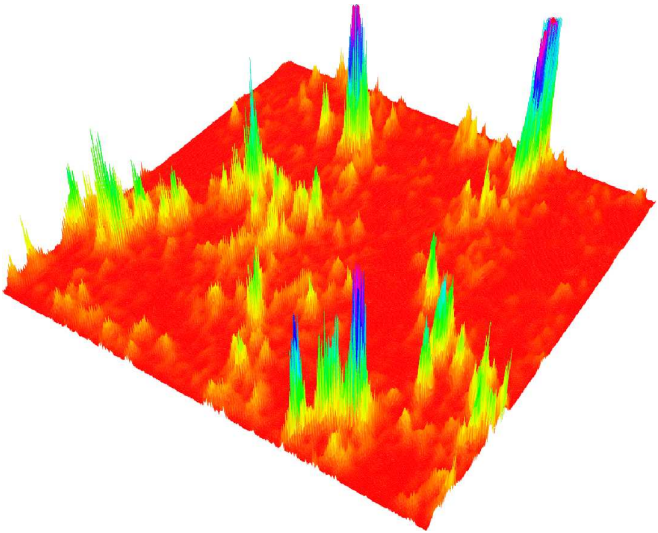}
\caption{(Color online) The probability density of a characteristic critical 
  quantum Hall eigenstate of the $n=0$ Landau band at $E=0.016815$. The system
  size is $L=320$, constant magnetic field is $p/q=1/32$, and random magnetic
  field strength is $f=0.1$.   
}
\label{Psi2}
\end{figure}

A similar behavior is found for the $n=0$ critical quantum Hall states, as well. 
This is shown in Fig.~\ref{Psi2} where the squared amplitude of a
characteristic eigenstate with energy $E=0.016815$ is displayed for a $L=320$
sample with $p/q=1/32$ and RMF strength $f=0.1$. A closer inspection reveals an
approximate sub-lattice polarization, also in this case. This effect appears to
be stronger than in the diagonal disordered situation.\cite{PS08}

The multifractal analysis was carried out as usual by utilizing the well-known
box counting method where the scaling of a ``box-probability'' is calculated,
$P(s,\lambda)=\sum_i^{N(l)}(\sum_{r\in\Omega_i(l)}|\psi_E(r)|^2)^s
\sim \lambda^{\tau(s)}$, from which the
generalized fractal dimensions $D(s)=\tau(s)/(s-1)$ or,
by a Legendre transform, the so-called $f[\alpha(s)]$ distribution
can be derived.\cite{Hea86,CJ89} Here, $\Omega_i(l)$ is the
$i$th box of size $l=\lambda L$ from which the $s$th moment of the
modulus of the normalized eigenstate $\psi_E(r)$ is taken.

\begin{figure}
\includegraphics[width=7.5cm]{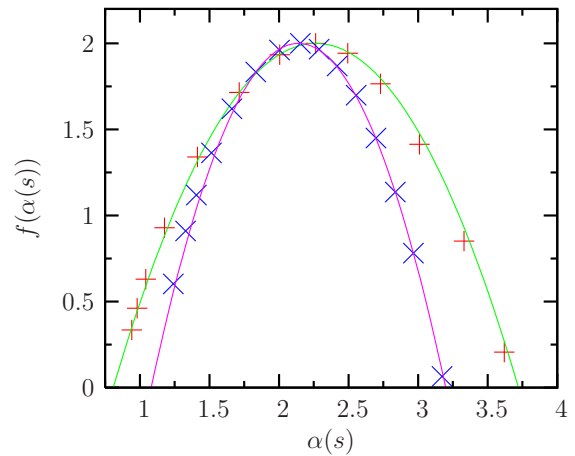}
\caption{(Color online) The $f[\alpha(s)]$ distributions for two chiral
  eigenstates with energy close to $E=0$. The parabolas are determined by 
  the values $\alpha(0)=2.14$ in the case of $B=0$, $f=1.0$ ($\times$), and
  $\alpha(0)=2.265$ for $p/q=1/32$, $f=0.5$ (+), respectively. In both cases,
  the system size is $L=320$. The data points correspond to $s=\pm 4.0$, $\pm
  3.0$, $\pm 2.5$, $\pm 2.0$, $\pm 1.5$, $\pm 1.0$, $\pm 0.5$, and 0.0. 
}
\label{falfa}
\end{figure}

In Fig.~\ref{falfa} we show the $f[\alpha(s)]$ distribution of two $d=2$ chiral
eigenstates at $E\approx 0$ in comparison with the parabolic approximation, 
$f[\alpha(s)]=d-[\alpha(s)-\alpha(0)]^2/\{4[\alpha(0)-d]\}$, which for finite
systems is normally valid only for small $|s|$. We found a value
$\alpha(0)=2.14\pm 0.02$ when averaged over several critical chiral states in
the case $B=0$. This value turns out to be close to the one obtained
previously for a square lattice with correlated random magnetic
field.\cite{PS99} For finite $B$, we found an averaged $\alpha(0)=2.27\pm 0.02$
which is the same as the $\alpha(0)$ value of the $n=0$ quantum Hall
states. This result also agrees with $\alpha(0)$ values published for various
quantum Hall models in the range $2.26 < \alpha(0) \le
2.29$.\cite{HKS92,HS94,KM95,EMM01} 
For the $n=1$ quantum Hall state the achievable system sizes were not
sufficient for the calculation of a conclusive multifractal spectrum. 

We finally mention that we also checked the $E=2.9725$ critical eigenstates
within the Landau band closest to the tight-binding band edge in the presence
of diagonal disorder. These outer states do not belong to the Dirac fermion
sequence that can be found only in the energy range $-1.0 \le E \le 1.0$. Again, we
obtained the ordinary quantum Hall multifractality with $\alpha(0)\simeq
2.29$ and a critical conductance $g_c\simeq 1.0\,e^2/h$ (including the aspect
ratio factor of 2).    

\section{Conclusions}
We investigated numerically a two-dimensional bricklayer lattice model, which
shares the same topology as graphene's hexagonal lattice, in the presence of
both a homogeneous and a spatially fluctuating random magnetic field. We
calculated the one-particle density of states, the two-terminal conductance,
and the multifractal properties of critical eigenstates. Within the very
narrow $n=0$ Landau band, we found a splitting of the two-terminal conductance
peak into three subpeaks. A central chiral conductance peak located at the
Dirac point $E=0$ with a scale independent value $4/\pi\,e^2/h \lesssim
g_c(E=0,f) \lesssim 8/\pi\,e^2/h$ (including the aspect ratio factor of 2), 
depending on the strength of the random magnetic field $f$.   
The symmetric splitting of the two other peaks $\Delta E(p/q,f)$ increases
with the square root of the applied perpendicular magnetic field expressed by
the rational number $p/q$ times a flux quantum $h/e$ per plaquette, and
linearly with the amplitude of the random magnetic-flux disorder $f$. The
scale independent conductance peak value of these critical quantum Hall states
is $g_c\approx 1.12\,e^2/h$ (including the
aspect ratio factor of 2), independent of disorder strength and applied
magnetic field. The splitting of the $n=0$ conductance peak allows for a
Hall plateau with both $\sigma_{xy}=0$ and $\sigma_{xx}=0$, except at $E=0$,
where the longitudinal conductance is finite due to the chiral critical state.  

A similar scenario, a splitting of the Landau band and an additional central
critical state with a $\ln(|E|^2$ singularity in the density of state  was
previously reported for a two state Landau model by Minakuchi and
Hikami\cite{MH96}. They found divergences of the localization length with an
exponent $\nu=0.26$ for the central state, which is not far away from our
result for the  chiral critical state, and for the split Landau band
$\nu=3.1$, which is somewhat larger but still believed by these authors to be
compatible with the conventional quantum Hall universality class. Presumably,
the large value has to be attributed to finite-size effects.    

Our analysis of the critical eigenstates revealed the sublattice polarization,
known from the clean system at $E=0$, to exist at least approximately also in
the random magnetic-flux disordered system for energies in the vicinity of the
Dirac point. Within the achieved uncertainty, the multifractal properties of
the critical eigenstates in the $n=0$ Landau band appeared to be the
same. We found the quantum Hall critical states parabolic $f[\alpha(s)]$
distribution, which is determined by an $\alpha(0)=2.27\pm 0.02$ value,  
to be in accordance with the multifractal properties of the chiral critical 
state close to the Dirac point.

\section{Acknowledgments}
We would like to thank Walter Apel and Ferdinand Evers for valuable discussions.
P.M. thanks Grant APVV Project No.~51-003505, VEGA Project No.~2/6069/26,
and PTB for hospitality.


\end{document}